\documentclass[preprint, nofootinbib]{revtex4-2}

\usepackage[dvipsnames]{xcolor}
\usepackage{physics}
\usepackage{ulem}
\usepackage{amsmath,amssymb,amsfonts,upref,endnotes}
\usepackage{amsthm}
\usepackage{crop}%
\usepackage{marginnote}
\usepackage{soul}
\RequirePackage{graphicx}
\usepackage[figuresright]{rotating}
\usepackage{color}

\begin{document}

\title{From quantum feature maps to quantum reservoir
computing: perspectives and applications}

\author{
Casper Gyurik\textsuperscript{1, \textdagger}, 
Filip Wudarski\textsuperscript{2, \textdagger}, 
Evan Philip\textsuperscript{1},
Antonio Sannia\textsuperscript{3,2},
Hossein Sadeghi\textsuperscript{1},
Oleksandr Kyriienko\textsuperscript{4} ,
Davide Venturelli\textsuperscript{2} ,
and Antonio A. Gentile\textsuperscript{1}
}

\address{\textsuperscript{1}Pasqal SaS, France
\\\textsuperscript{2}USRA Research Institute for Advanced Computer Science (RIACS), USA
\\\textsuperscript{3} Institute for Cross-Disciplinary Physics and Complex Systems (IFISC) UIB-CSIC, Campus Universitat Illes Balears, 07122, Palma de Mallorca, Spain
\\\textsuperscript{4}School of Mathematical and Physical Sciences, University of Sheffield, UK 
\\\textsuperscript{\textdagger}\textit{These authors contributed equally}
}

\begin{abstract}
We explore the interplay between two emerging paradigms: reservoir computing and quantum computing. 
We observe how quantum systems featuring beyond-classical correlations and vast computational spaces can serve as non-trivial, experimentally viable reservoirs for typical tasks in machine learning.  
With a focus on neutral atom quantum processing units, we describe and exemplify a novel quantum reservoir computing  (QRC) workflow. 
We conclude exploratively discussing the main challenges ahead, whilst arguing how QRC can offer a natural candidate to push forward reservoir computing applications.
\end{abstract}

\maketitle

\section{Introduction}

Quantum computing (QC) is a specific model of computation that uses the principles of quantum mechanics to process information. 
Unlike the classical bits (0 or 1) used in the von Neumann architecture, quantum computers use qubits, allowing for superposition states.
Quantum systems exhibit interference and entanglement, which allow qubits to be correlated in ways that classical bits cannot. 
Digital computation in this model is carried out using quantum gates, which are operations that change the state of qubits, similar to how logic gates operate on bits in classical circuits. 
Analog operations where the qubits are evolved according to their intrinsic system behaviour are also possible in quantum circuits.

While quantum computers grow in size, they remain noisy and limited in size, with a number of qubits growing to hundreds, they are still sparsely connected and have non-negligible error rates~\cite{proctor2025benchmarking}. Hybrid quantum-classical approaches offer the possibility of enhancing capabilities of quantum devices and tailoring them for specific applications~\cite{callison2022hybrid, wudarski2023hybrid}. 
These algorithms show modularity and flexibility, allowing them to exploit existing hardware while delivering the best performance. Additionally, one can scale the size of the system in a modular way, connecting registers with a handful of qubits into correlated multi-register systems with hundreds. 
The former can serve as a test-bed for theoretical investigations, as well as numerical simulations, that could be used as a performance benchmark for quantum hardware runs. 

Quantum computing can be seen as a part of a broader field of physics-based computing, encompassing also analog, reservoir and neuromorphic frameworks. 
In particular, reservoir computing leverages upon a complex dynamical system to map inputs into a high-dimensional computational space, and processes readout data to efficiently solve learning tasks, including time-series predictions and classification~\cite{platt2022systematic, yan:qrc_overview}. 
Neuromorphic computing explores novel ways to process information, attempting to mimic brain processes with the intent of matching its efficiency. In this regard, some routes towards neuromorphic computing entertain the adoption of reservoir-like mappings enabled by biological or physical systems~\cite{markovic2020neurophysics}. Reservoir computing approaches exemplify heuristic computational paradigms. These methods typically lack formal performance guarantees but rely on empirical validation, offering solutions to complex or high-dimensional tasks that otherwise are prohibitively expensive~\cite{gauthier2021nextrc}. 

The main appeal of reservoir computing is twofold: (i) because only the linear readout is trained, the models are relatively easy to optimize while still potentially capturing temporal dependencies (unlike recurrent neural networks, which are notoriously difficult to train due to challenges in backpropagation); and (ii) with appropriately chosen reservoirs, both the computational and energy costs of training and inference can be much lower compared to other approaches~\cite{gauthier2021nextrc, yan:qrc_overview}. 
The promise of QC resides in the favourable scaling of computational resources, compared to the best-known classical methods, expected when quantum algorithms are applied to certain problems, such as factoring large numbers or simulating quantum systems, making QC an obvious candidate in the efforts to optimise the computing load across several disciplines~\cite{huang2025vast}. 

In this contribution, we explore quantum reservoir computing (QRC) as an emerging machine learning paradigm that combines quantum evolution with non-trivial, high-dimensional feature processing.
We concentrate on physical systems that are based on neutral atom arrays, with arbitrary connectivity and strong interactions.
Such systems can naturally simulate complex multi-qubit models and their dynamics, satisfying one of the main requirements of reservoir computing.
We describe in full generality the QRC paradigm, and highlight the differences from other Quantum Machine Learning (QML) approaches. 
We complement such analysis with exemplary applications of neutral atoms to relevant tasks in QRC, suggesting a novel hybrid scheme to leverage NISQ devices for early implementations.

Building on such discussion, we argue how QC can amplify other non-von Neumann models of computation, providing benefits from coherent operation and entanglement, and in particular offer a promising venue to expand efforts in the direction of reservoir and neuromorphic computing paradigms.

\section{Basics of reservoir computing}
\label{sec:RC_basics}

Reservoir computing (RC) systems have been long proposed and successfully applied across a wide range of domains. 
As outlined in~\cite{yan:qrc_overview}, these applications can be broadly categorized into a few main areas: signal classification (e.g., spoken digit recognition), time series prediction (e.g., stock market forecasting), control of dynamical systems (e.g., real-time robot control), and PDE-based computations (e.g., fast simulation of certain physical systems governed by differential equations).

Formally, a RC system operates as follows.  
Let the input data be denoted by $u_1, \dots, u_t$, where each $ u_i \in \mathbb{R}^n $, and let $ u_i(j) $ denote the $ j $th entry of $ u_i $.  
We index the data according to the order in which it is presented to the learner, to emphasize potential temporal dependencies.
The dynamical map that describes the reservoir evolution is denoted by $ \mathcal{R} $, and in typical digital implementations, it is parameterized by a set of (fixed) random matrices $W$. At each time step $t$, the reservoir state $ r_t $ is given by the recursive relation
\begin{equation}
r_t = \mathcal{R}(u_t, r_{t-1};W) \in \mathbb{R}^N.
\label{eq:reservoir_state}
\end{equation}
For a suitable choice of $\mathcal{R}$, $r_t$ is a function of the past inputs up to a typical characteristic time scale, $t_{\mathrm{FM}}$, known as fading memory time, e.g.
\begin{equation}\label{Eq:fading_memory}
    r_t = r_t(\{u_i\}_{t-t_{\mathrm{FM}}}^t),
\end{equation}
allowing the system to encode temporal dependencies in its state.

Finally, a linear readout layer $ W_{\mathrm{out}} \in \mathbb{R}^{m \times N} $ is trained to produce  
\begin{equation}
y_t = W_{\mathrm{out}} r_t,    
\label{eq:readout}
\end{equation}
where $ y_t \in \mathbb{R}^m $ is the output, i.e. the  reservoir prediction at time $ t $. Therefore, we can distinguish three main subroutines: i) data embedding, ii) dynamical mapping, and iii) post-processing for training and prediction.
When studying what makes a good reservoir, there are certain key considerations. 

A crucial one is whether the reservoir exhibits some form of \textit{fading memory}.
This requirement is twofold: first, the state of the reservoir should depend on past inputs. However, the reservoir should primarily depend on {recent} inputs rather than inputs from the distant past. 
The first aspect is typically quantified by the \textit{short-term memory capacity} of the system~\cite{jaeger:memory}, while the second is referred to as the \textit{echo state property}~\cite{jaeger:echo} and can be measured, for instance, using the {echo state property index}~\cite{gallicchio:echo_measure}. 
The fading memory is imprinted naturally in the dynamical map Eq.~\eqref{Eq:fading_memory}, which exploits non-linear activation functions crucial for good performance.

Equally important is the presence of \textit{non-linearity} in the reservoir internal dynamics. Since the readout layer is typically linear like in Eq.~\eqref{eq:readout}, the reservoir must introduce sufficient complexity to allow for non-trivial transformations of the input. This often requires tuning the system close to the {edge of chaos} --- a regime where the dynamics are rich and sensitive to inputs, but not so unstable that they become unpredictable, or input information is completely scrambled. This balance ensures a high-dimensional, expressive {training} space while keeping enough structure for meaningful processing.

A significant portion of research in RC is dedicated to identifying physical systems that can effectively implement the reservoir mapping $r$ specified in Eq.~\eqref{eq:reservoir_state}. 
As detailed in Ref.~\cite{yan:qrc_overview}, various platforms have been explored for this purpose, including but not limited to: memristors~\cite{zhong:memristors}, spintronic devices~\cite{finocchio:spintronics}, photonic systems~\cite{van:photonics}, and digital FPGAs and ASICs~\cite{penkovsky:FPGA}.

\section{Quantum systems as reservoirs}
\label{sec:intro_QRC}

A novel proposal in this direction is to use quantum systems -- particularly quantum computers, whether digital or analog -- as reservoirs.
Quantum systems are {solid} candidates for reservoirs due to their inherently complex, high-dimensional state spaces and non-trivial dynamics. 
The underlying idea is straightforward: quantum states live in exponentially large Hilbert spaces, making them natural {candidates} for generating high-dimensional representations suitable for training a linear readout. 
Inspired by features emerging from the QML field -- as discussed in Sect.~\ref{sec:fm_to_reservoir} -- a growing body of literature has investigated how to effectively leverage quantum systems as reservoirs, as pioneered by works such as \cite{fujii:qrc_og, markovic:qrc_og} and surveyed e.g. in \cite{mujal:qrc_overview}.

In this contribution, we emphasise how it can be especially promising to adopt quantum systems exhibiting strong interactions such as lattices of Rydberg atoms (qubits), see Sect.~\ref{sec:NA_intro}. 
These systems naturally support rich entanglement and strongly-correlated behaviour, allowing them to efficiently encode and process temporal correlations in the inputs. 
Moreover, their tunable interaction strengths and coherence properties enable precise control over the dynamical regime, making it possible to steer the system toward the optimal balance between memory and nonlinearity required for effective reservoir computing.

\subsection{From quantum feature maps to quantum reservoirs}
\label{sec:fm_to_reservoir}

In classical machine learning, \textit{feature mapping} (FM) is a well established concept whereby a certain vector of data $u \in \mathcal{U}$ (the feature)\footnote{For notation brevity we use $u$ for both vectors and scalars, and the nature of $u$ is self-explanatory from the context.} 
gets mapped via a transformation $f$ from the initial space $\mathcal{U}$ to a different feature one, in order to better exploit some properties of the transformed dataset $ f(u) \in \mathcal{F}$. 
A commonly used example is achieving the linear separability of a dataset for classification purposes, when adopting a Support Vector Machine (e.g. \cite{albrecht2023quantum}).

A key problem in QML is to load a typically classical input on a quantum device, in order to further process it.
The way to achieve this {typically} invokes a \textit{quantum feature map} acting on a (classical) input $u$, and targeting a quantum Hilbert space $\mathcal{H}$ such that:
\begin{equation}
    u \in \mathcal{U} \rightarrow \ket{\psi(u)}.
    \label{eq:fm}
\end{equation}
This mapping is often a quantum circuit parameterised via $u$ known as the \textit{ansatz}, and acting as a unitary transformation $U_{\psi}[f(u)]$ of an initial fiducial state, onto the quantum state $\ket{\psi(u)} \in \mathcal{H}$, where we adopted the common \textit{braket} notation. 
The features $u$ can then be extracted (in part or in full) by measuring the expectation value of an observable as 
\begin{equation}
\langle O\rangle =\bra{\psi(u)}O\ket{\psi(u)},     
\label{eq:observable}
\end{equation}
which implies an additional (non-linear) transformation of the data. 
Therefore, one may treat not only the data encoding $f$ but also the associated measurement as the feature map, as the latter is inherited in order to extract any information from a quantum state.
Various possible encodings are possible, impacting the scaling in the width and depth of the quantum circuit required to embed a dataset of given dimensionality, e.g. $M$ copies of size $N \equiv \dim(u)$, but also the suitability of the protocol in near-term quantum devices. 
A review of the main possible techniques can be found e.g. in \cite{schuld2019quantum}, and state-of-art methods have produced highly expressive feature maps by appropriately combining classical and quantum operations~\cite{kyriienko2021solving, albrecht2023quantum}.

\begin{figure}
    \centering
    \includegraphics[width=0.32\linewidth]{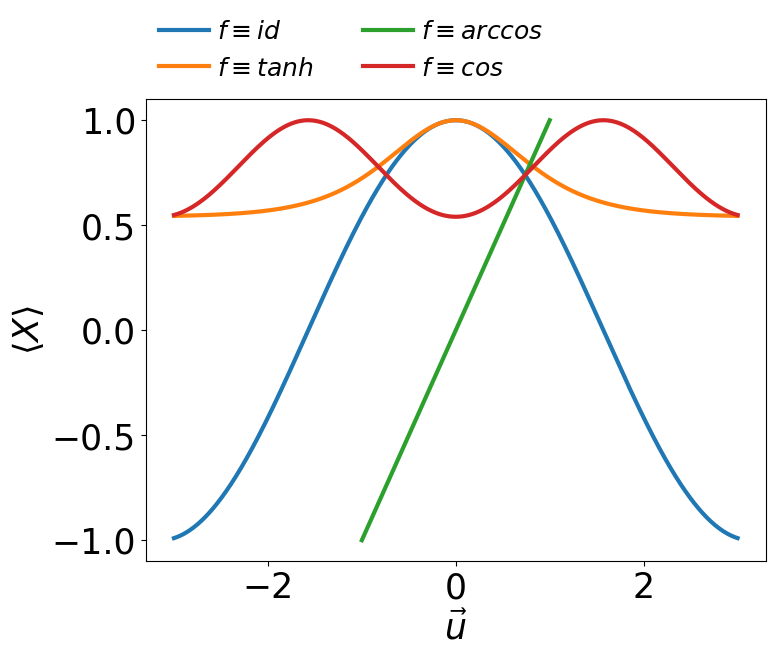}
    \includegraphics[width=0.32\linewidth]{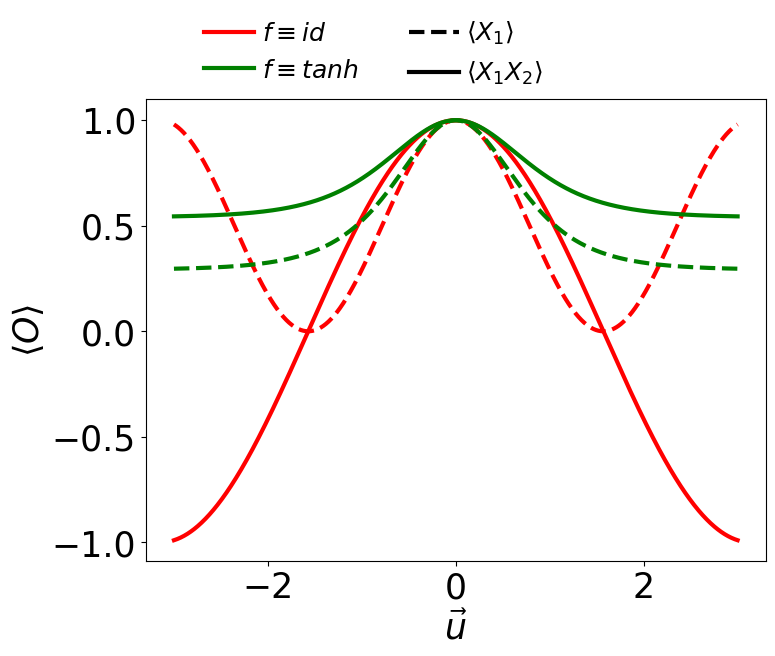}
    \includegraphics[width=0.34\linewidth]{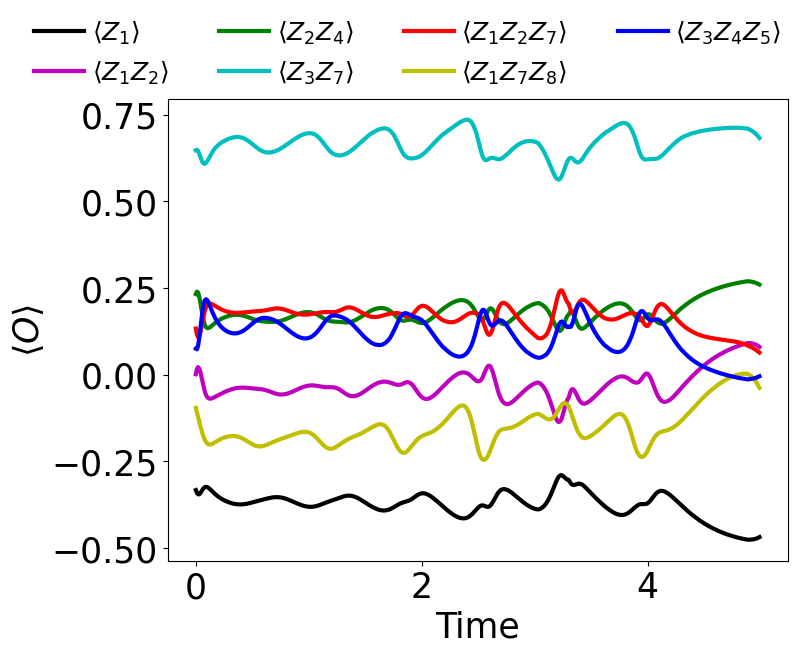}
    \caption{Impact of various feature map encodings. 
    (Left) Single-qubit case with 1D data $u$ on the outcome, expressed as the expectation value of a state $R_Z[f(u)]\ket+$. 
    (Middle) 2-qubit case, where two types of observables are extracted, as distinguished by dashed versus solid lines. The colours identify different feature map encodings (identity and $\tanh$). 
    (Right) 8-qubit case encoding a 3D time-series, i.e. $\dim(u)=3$, and values of various colour-encoded observables as a function of time.}
    \label{fig:feature_maps}
\end{figure}

We illustrate various feature maps (both on the level of unitary encoding and measurement) in Fig.~\ref{fig:feature_maps}. 
First, we consider the most basic scheme of a single-qubit system that is transformed with respect to a single-qubit rotation $R_Z(\alpha) = \exp(-i \alpha Z/2 )$, where $Z$ is the Pauli-$Z$ matrix, and the initial state $\ket{+} = 1/\sqrt{2}(\ket{0}+\ket{1})$. 
Taking Pauli-$X$ as the observable $O$, the expectation value with respect to the output state $\ket{\psi(a)}=R_Z(a)\ket{+}$ is given as $\langle X\rangle= \cos(a)$. 
Note that the cosine argument can also undergo a feature map transformation as $a\equiv f(u)$, yielding $\langle O\rangle = \cos[f(u)]$. We illustrate also the impact of replacing the identity mapping $f\equiv id$ with various non-linear functions: $f\equiv \tanh$, $f \equiv cos$ and $f\equiv arccos$\footnote{Note that in case of $\arccos(\cdot)$ we operate on a truncated domain}. 
Similarly in Fig.~\ref{fig:feature_maps}(middle), we consider two-qubit case, where the initial state $\ket+\ket+$ is first rotated with $R_Z(a)\otimes R_Z(a)$ (same $Z$ rotation acting on each qubit) followed by $CX$ gate. In 2-qubit case, we have more abundant structure of observables that one can measure, and we compare single-qubit expectation value $\langle X_1\rangle$ and 2-qubit correlator $\langle X_1X_2\rangle$ with different feature maps encoding the input data. For both single- and two-qubit toy examples, one notices that different encodings transform the data in completely different manner, leaving us with a flexible subroutine to meet machine learning objectives. 
Finally, for a multiqubit system and multi-parameter input data (previously $u\equiv u$ was a scalar), the landscape looks more complex. 
To illustrate this, we plot in Fig.~\ref{fig:feature_maps}(right) various observables extracted from encoding input data from the chaotic system Lorenz63
detailed and adopted later in Sec.~\ref{sec:hybrid_framework}. 
Here, the dependency is against time, as the Lorenz63 is a 3D time-series and the encoding phase mixes all components of the input vector together. 

When bearing in mind the requirements for a good reservoir structure elicited in Sect.~\ref{sec:intro_QRC}, a first question concerning Eq.~\eqref{eq:fm} is: to what extent can we introduce non-linearities in a quantum reservoir adopting a FM scheme?
In the classical case, such non-linearities where naturally present in the dynamical map.
However, the evolution of a closed quantum system (an approximation valid for many quantum devices) is linear and described by unitary transformations $U$. Then, a natural choice is to invoke a non-linear behaviour within the feature map encoding of the data. That is, $f(u)$ shall be a non-linear function of $u$.

Additional challenges and opportunities arise when considering the way we extract classical information from a quantum system -- i.e. the measurement. The latter invokes the estimation of \textit{observables}, i.e. special operators acting on the system's Hilbert space.
In a quantum reservoir framework, this measurement step preludes the final, linear readout operation in Eq.~\eqref{eq:readout}.
However, a well-known aspect of quantum measurement is that it collapses the internal state of the system in a non-reversible fashion, producing a classically processable \textit{bitstring} as a result, of the same size as the number of qubits in the device. Each bit therein represents the collapsed value in the computational basis of the corresponding qubit, and average estimates like Eq.~\eqref{eq:observable} require access to many copies of the same state, and repeated measurements. 

The requirements to build a reservoir are somewhat different from other QML tasks like linear regression or classification, where the reset of the system's memory after each measurement does not pose any fundamental issues. 
Fading memory requirements conflict with ``forgetting'' the process generating the output quantum state - which is, to some extent, the outcome of the measurement process. 
Therefore, other strategies are required to restore the ability of a quantum system to retain memory of past inputs, highlighting a central challenge in quantum reservoir computing design, that we discuss in further details in Sect.~\ref{sec:outlook}.

\subsection{Neutral atom dynamics}
\label{sec:NA_intro}

Let us consider a specific example of a quantum system suitable for use as a reservoir: analog quantum simulators based on neutral atoms, particularly arrays of Rydberg atoms~\cite{henriet:pasqal}.
These systems are currently capable of storing hundreds of physical qubits in vacuum chambers sitting at room temperatures, which can be manipulated by laser addressing of appropriate frequency, and engineered pulses. 
In general, such a system comprising $n$ atoms can be described by a Hamiltonian of the form\footnote{Hereby we assume $\hbar=1$ to simplify notation.}:  
\begin{equation}
H(\{\Omega_j\}, \{\delta_j\}, U, t) = \sum_{j=1}^n \left( \Omega_j(t)\,X_j + \delta_j(t)\,n_j \right) + \sum_{i < j} U_{ij} (t) n_i n_j,
\label{eq:NA}    
\end{equation}
where  $n_j = (1 + Z_j)/2$ is the number operator.  
The first two parameters $\Omega$ (\textit{amplitude}) and $\delta$ (\textit{detuning}) are connected with the laser light - atoms interaction, whereas the \textit{coupling} strengths $U_{ij}$ can be tuned independently from each other. In principle, all three parameters are dependent on time $t$, as pulses can be modulated, and the atom positions can be modified via optical-tweezers during the computation allowing for additional operations~\cite{bluvstein2024logical}.

{In our tests we} tend to implement less general versions of Eq.~\eqref{eq:NA}. 
The three parameters can often be considered static in time, as some Neutral-Atom devices have adopted fixed atom arrays throughout the computation (thus making $U_{ij}$ constant), and the modulation of the pulses occurs on timescales fast enough, that can be considered constant at least block-wise in terms of overall system evolution. 
An important simplification is also attained by adopting \textit{global} $\Omega, \delta$ for the whole array, as this removes the demanding requirement of laser-addressing single atomic sites independently~\cite{henriet:pasqal}.
When including all such simplifications, we are left with a type of Transverse Field Ising Model (TFIM) Hamiltonian~\cite{labuhn2015realizing}:
\begin{equation}
H(\Omega, \delta, U) = \sum_{j=1}^n \left( \Omega\,X_j + \delta\,n_j \right) + \sum_{i < j} U_{ij} n_i n_j.
\label{eq:Ising}    
\end{equation}
Beside the obvious potential of directly quantum-simulating such models directly on a neutral atomic system, Eq.~\eqref{eq:Ising} can still be expressive enough to implement machine-learnable models, as demonstrated e.g. in \cite{albrecht2023quantum}.

\subsection{Hybrid Quantum-Classical Reservoir Computing with Neutral Atoms}
\label{sec:hybrid_framework}

In the spirit of targeting early experimental implementations on NISQ available quantum devices, we focus now on describing a possible \textit{hybrid} QCRC (hQCRC) framework, suitable in particular to the neutral atom platform introduced in Sect.~\ref{sec:NA_intro}. 
A hybrid approach can serve as a remedy for fading memory~\cite{kornjavca2024qrc}, as (in its most elementary design) the dynamical map can be split into two parts: classical and quantum. The former handles memory effects, while the latter exploits an exponentially large Hilbert space for the data encoding. Both parts process their respective data streams with non-linear characteristics, thus having the necessary components to form a good RC architecture.

\begin{figure}
    \centering
    \includegraphics[width=0.95\textwidth]{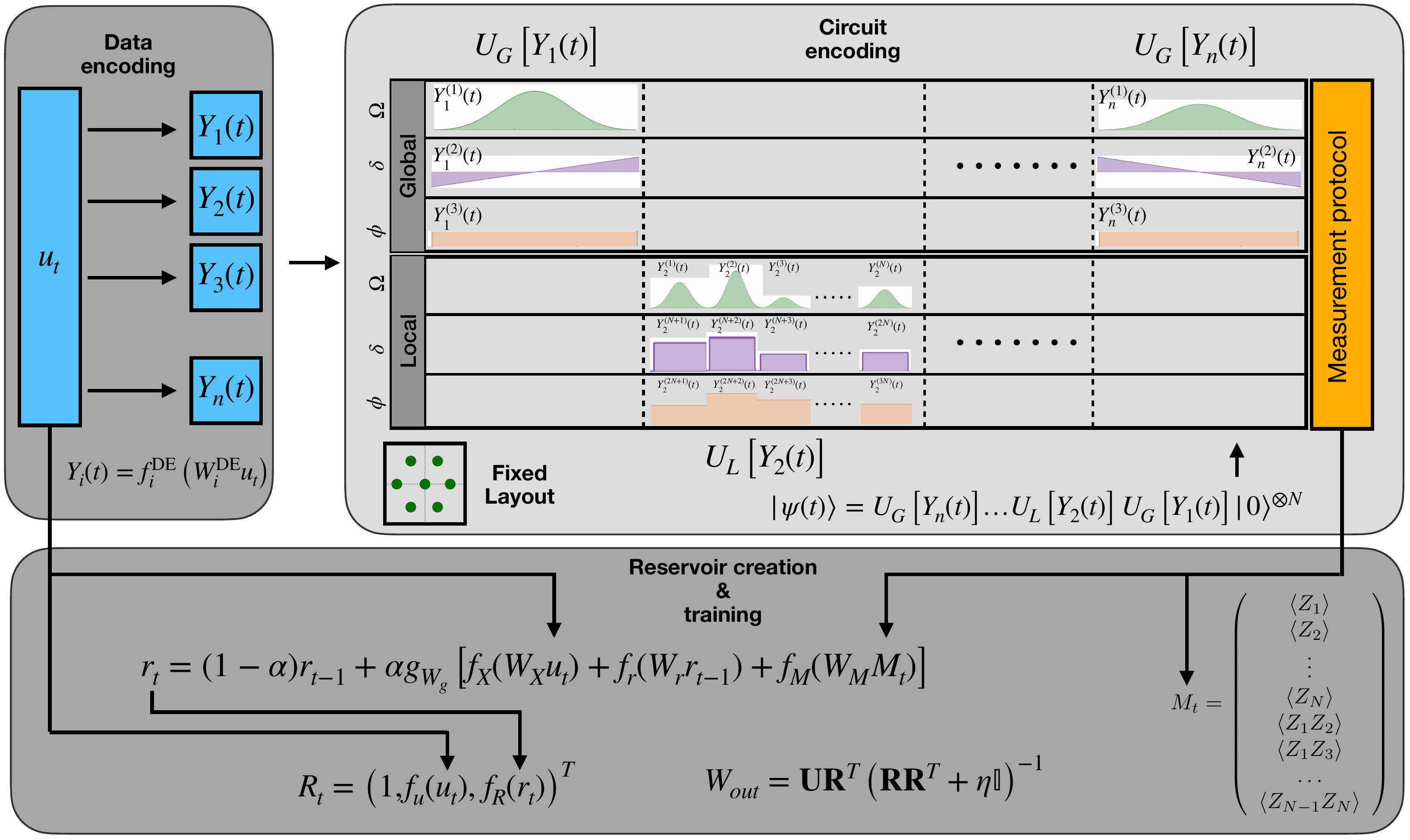}
    \caption{
    A representation of the 3 phases of the suggested hybrid Quantum Reservoir Computing protocol. 
    A data encoding manipulates the initial vector $u_t$ into a modified $Y_i(t)$ which is then encoded into the quantum circuit, where $i$ indicates the layer in which the data is injected. The encoding is here exemplified for a Neutral Atom architecture compatible with Eq.~\eqref{eq:NA}, and processes the input $Y_i^{(j)}(t)$ {according to the chosen (sets of) free parameters $j$ such as $\Omega, \delta, \phi$}, alternatively operating at a global or local level (details in the main text). The pulse sequences shown for the circuit encoding are generated in \texttt{Pulser}~\cite{silverio2022pulser}.
    The readout from the quantum circuit $M_t$ is finally post-processed in a reservoir-like fashion compatible with the relation \eqref{eq:reservoir_state}, and trained until $r_t$ reproduces the target dataset. 
    Phases implemented in classical (quantum) devices are encircled in dark (light) gray boxes. 
    }
    \label{fig:hQCRC_network}
\end{figure}

The framework we propose is inspired by \cite{wudarski2023hybrid} (see also \cite{settino2025memory} which further exploits classical processing techniques), with adjustments to the neutral atom platform, which will eventually provide digital-analog capabilities as in \cite{parra2020digital}. A schematic workflow focusing on sequential datasets is depicted in Fig.~\ref{fig:hQCRC_network}. 
The framework operates in three phases: 
\begin{enumerate}
    \item classical data pre-processing --- responsible for creating suitable input data to feed in the dynamical map as mentioned in Sect.~\ref{sec:RC_basics}, 
    \item quantum encoding --- a subroutine of the dynamical mapping whose general behaviour we described in Eq.~\eqref{Eq:fading_memory}, 
    \item classical post-processing --- combining contributions from the classical and quantum dynamical maps, whose output serves training or prediction purposes. 
\end{enumerate}
This is evidently a hybrid classical-quantum-classical scheme, relying on various transformations. 
First (i), the input data is prepared as $\{Y_i(t)\}$ to fit the quantum \textit{ansatz}, representing a quantum subroutine of the dynamical map, 
hence it is transformed via combination of linear and non-linear transformations $f_i^{\mathrm{DE}}\colon D_{\mathrm{input}}\to D_{\mathrm{layer}}$, where $D_{\mathrm{input}/\mathrm{layer}}$ is the dimension of input/layer. 
Linear transformations are matrix multiplication with fixed - usually random - matrices $W_i^{\mathrm{DE}}$ -- see also Eq.~\eqref{eq:readout} -- whereas the non-linear ones are activation functions similar to those elicited in Fig.~\ref{fig:feature_maps}.

The input data is transformed $n$-times with different (random) $W_i^{\mathrm{DE}}$ and different $f_i^{\mathrm{DE}}$, where $n$ is the number of parameterized (data-dependent) layers in the quantum circuit (ii). 
The quantum ansatz is constructed by first fixing the arrangement of atoms\footnote{
In principle the arrangement could also depend on transformed data $Y_i(t)$, however changing the arrangement from iteration to iteration extends {wall-clock} execution time, therefore one needs to take these overheads into account.}, 
which is responsible for interaction between atoms and thus also the generation of entanglement.    
Next a series of laser pulses, algorithmically arranged into layers, are applied to encode the transformed data into a quantum state through time-dependent evolution. 
Each layer has a number of tunable parameters: $\Omega$, $\delta$, $\phi$ (described in Sect.~\ref{sec:NA_intro}) as well as the duration (which is fixed per-layer in our experiments).
The number of terms depends on the type of pulses that are applied, e.g. with local pulses one has the flexibility to address $1$ to $N-1$ different atoms, while global pulses addressing the whole atom register allow for adjustment of only the introduced parameters.
Also, in principle the shape of each pulse can be modelled in a time-dependent fashion; however, constant pulses can still provide sufficient data encoding expressivity.
Future generations of neutral atom platforms are expected to provide even more flexibility, for example, Raman transitions and in-processing layout re-arrangement are areas of active research \cite{henriet:pasqal, bluvstein2024logical}. 

The quantum state after the full evolution, i.e. when all layers have been applied, has the form
\begin{equation}
    |\psi(t)\rangle = U_G\left[Y_n(t)\right]\ldots U_L\left[Y_2(t)\right]U_G\left[Y_1(t)\right]|0\rangle^{\otimes N}\ ,
    \label{eq:qrc_state}
\end{equation}
where $U_{G/L}[Y_i(t)]$ is a global/local unitary (pulse) parameterized by transformed data $Y_i(t)$. This state is then measured according to a measurement protocol that enables us to create a measurement vector $M_t$ composed of expectation values of selected observables. For example, for $Z$ type measurements, one may create a measurement vector comprised of single-qubit expectation values and two-qubit correlators as
\begin{equation}
    M_t = \left( \langle Z_1\rangle, \langle Z_2\rangle\ldots,\langle Z_N\rangle \langle Z_1Z_2\rangle, \ldots \langle Z_{N-1}Z_{N}\rangle\right)^T\ ,
    \label{eq:expvals}
\end{equation}
where the ansatz design specify which qubits and what correlators are used. 
The goal of a quantum encoding subroutine is to spread classical data in a high-dimensional Hilbert space, {whilst preserving the possibility for measurement protocols to extract such compressed information.} 
Hence, the structure of the protocol is treated as a hyperparameter of the network that needs to be carefully selected, ideally taking into account trade-offs between the number of the measurements and informational content they carry on in a measurement vector.

As a final step (iii), we process available streams of data: input data, measurement vector and previous reservoir states to construct a compact reservoir state $r_t$
\begin{equation}\label{Eq:ESN}
    r_t = (1-\alpha) r_{t-1} + \alpha g_{W_g}\left[f_X(W_X u_t) + f_r (W_r r_{t-1}) + f_M(W_M M_t)\right]\ ,
\end{equation}
where $\alpha$ is a leak rate, $f_X, f_r, f_M$ are non-linear functions responsible for transforming the input data $u_t$, previous reservoir state $r_{t-1}$ and measurement vector $M_t$, that first undergo linear transformation with fixed (usually random) matrices $W_X, W_r$ and $W_M$, respectively. The entire second part can additionally be transformed in a similar way, i.e. a combination of linear and non-linear parts with a function $g_{W_g}\equiv g(W_g\cdot)$. The reservoir state update is responsible for memory effects by mixing previous states with current construction, allowing it to learn time characteristics. The reservoir states are constituents of a reservoir vector $R_t$
\begin{equation}
    R_t = \left(1, f_u(u_t), f_R(r_t)\right)^T\,
\end{equation}
where unity plays the role of a bias, and $f_x,f_R$ are non-linear activation functions for the input data $u_t$ and the reservoir state $r_t$\footnote{In principle, one can combine these non-linear transformations, with yet another set of linear transformations. Our numerical experiments on small system sizes suggest that it is unnecessary to obtain reasonably good outcomes.}, respectively. The reservoir vectors are then aggregated into a matrix $\mathbf{R}=[R_p, R_{p+1},\ldots, R_{d}]$, where we discarded the first $p-1$ vectors to avoid reliance on the initial conditions, and kept remaining vectors from the training set ($t=1,2,\ldots, d$, keeping $d-p$). The collected reservoir vectors are used to determine an output matrix $W_{out}$ in the ridge regression procedure that aims to find least-square mean between target data $\mathbf{u}=[u_p,u_{p+1},\ldots, u_d]$ and the training data $\mathbf{R}$ as
\begin{equation}
    W_{out} = \mathbf{U}\mathbf{R}^T\left(\mathbf{R}\mathbf{R}^T +\eta \mathbb{I}\right)^{-1},
\end{equation}
where $\eta$ is regularization coefficient that stabilize the solution. The output matrix is then used to predict future time-steps as
\begin{equation}
    y_{t+1} = W_{out} R_t\ , \quad \mathrm{for}\quad t=d, d+1,\ldots.
\end{equation}
This setup gives us a flexible and modular approach that can be easily adjusted to hardware restrictions, e.g. the currently cloud-available neutral-atom hardware supports {\it global} channels only~\cite{silverio2022pulser, pasqalcloud}. Additionally, it provides a great test-bed for investigating the transition between classical and quantum reservoir computing. In particular, one can turn off the quantum part by setting $f_M\equiv 0$ or turn off the dependence of the reservoir states on the input data (setting $f_X\equiv0$) to investigate the importance of quantum or classical contribution. The standard echo state network  is recovered for $f_X=f_r=id$, $f_M\equiv0$ and $g_{W_g}=\tanh$ (see, for example, \cite{lukovsevivcius2012practical}). 

\begin{figure}
    \centering
    \includegraphics[width=0.96\linewidth]{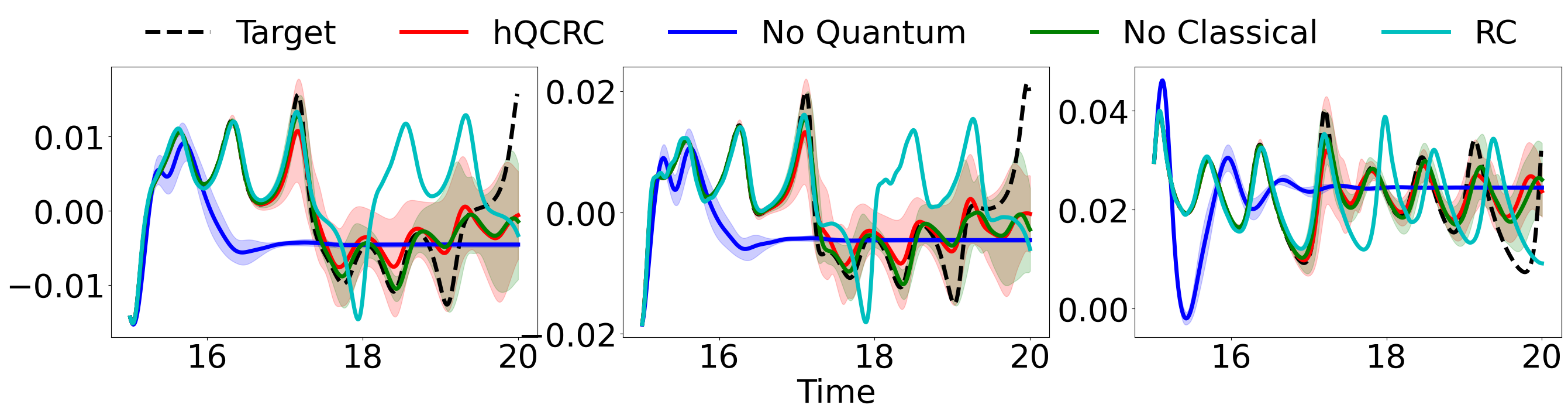}
    \caption{Prediction phase of the normalized Lorenz63 time-series benchmark (black) for $x, y$ and $z$ components (left to right). The red line represents a numerical simulation of the hQCRC framework in \eqref{Eq:ESN} (i.e. $f_M\equiv id$), whereas the blue line synthetically retrieves a classical-phase only performance by imposing $f_M\equiv0$ in the same equation, while the green line uses $f_X\equiv 0$, i.e. the reservoir $r_t$ depends only on measurements extracted from the quantum states $M_t$. {Finally, the teal line depicts the best performing setup for classical RC.
    Implementations details of the circuit in the main text}.
    The shaded region describes $\pm$ standard deviation over 20 random seeds of quantum reservoirs.}
    \label{fig:results}
\end{figure}

As a demonstration of the functionality of the approach, we present in Fig.~\ref{fig:results} a comparison between the predictions from Eq.\eqref{Eq:ESN} with and without the quantum part, i.e. when $f_M \equiv id$ and $f_M\equiv 0$, respectively. 
The results here are based on a standard chaotic benchmark - Lorenz63 \cite{lorenz1963deterministic,platt2022systematic} - that allows us to assess the performance of the framework in a challenging time-series forecasting task. 
For the quantum phase, we adopt an ansatz spanning an 8-qubit register, where the measurement protocol is fixed, composed of $Z$-type measurements to get $\langle Z_i\rangle, \langle Z_iZ_j\rangle$ and $\langle Z_iZ_jZ_k\rangle$ on all possible combinations of $i,j,k$. We stress that a fixed approach, may not necessarily be the most optimal, as it lacks any information or feedback about the problem. 
The purpose of the displayed results is to demonstrate the relevance of the quantum part, as all other parameters are the same, serving as a fair comparison between classical and quantum. Therefore, one may perceive the contribution from the quantum processing as a beneficial addition to the classical system, which is based on exploiting the Hilbert space (circuit) encoding and extracting of information from the quantum states via measurements. 
Note that we did not optimize over the ansatz hyperparameters, leaving this additional improvement for future contribution, and rather treat these results as a representative co-processing boost that can be attained by our hQRC approach, when running on an analog neutral atom platform. A more thorough benchmark will be required to assess average performances and benefits of our protocol (similar to what attained in \cite{wudarski2023hybrid} for a gate-based approach). 
Further details are reported in an external repository~\cite{zenodo}.

The purpose of the displayed results is to demonstrate the relevance of the quantum part, as all other parameters are the same, serving as a fair comparison between classical and quantum. Therefore, one may perceive the contribution from the quantum processing as a beneficial addition to the classical system {(blue line in Fig.~\ref{fig:results})}, which is based on exploiting the Hilbert space (circuit) encoding and extracting of information from the quantum states via measurements. 
Note that we did not optimize over the ansatz hyperparameters.
These results are thus intended as a representative co-processing boost attainable via our hQCRC approach on existing analog neutral atom platforms. More thorough benchmarks will be required to assess average performances and ultimate benefits of a hybrid protocol (as attained in \cite{wudarski2023hybrid} for digital approaches). 

\section{The road so far, the road to come}
\label{sec:outlook}

In Sect.~\ref{sec:intro_QRC}, we have elicited some fundamental challenges impacting in general the use of quantum systems as performing reservoirs beyond feature mapping -- such as the ability to retain fading memory over sequences of inputs, a hallmark of RC. 
We thus described in Sect.~\ref{sec:hybrid_framework} how one can defy such criticalities in an exemplary demonstration, focusing on a neutral atom platform and benefitting from a hybrid architecture to bypass some desirable reservoir properties lacking from what readily implementable in the described quantum device. 
Here, we intend to discuss in perspective some of these design choices, and how they might evolve in the near future. 

While quantum systems are naturally suited for generating expressive, high-dimensional representations, ensuring that they also possess the temporal dynamics necessary for memory remains non-trivial. We elicited before the incompatibility between projective quantum measurements and the necessity for the reservoir to retain memory during the processing. 
A straightforward workaround is to reinitialize and evolve the quantum reservoir from scratch for each new data point, but this quickly becomes intractable as the number of inputs grows.
Hybrid approaches like ours, or the one introduced in \cite{kornjavca2024qrc}, outsource the memory build-up to classical co-processing. 
A more advanced approach for mitigating this issue would be to adopt so-called \textit{weak measurements}, which are agnostic to the specific quantum hardware used. 
Rather than performing a full projective measurement -- as illustrated before, which collapses the quantum state entirely -- weak measurements couple ``reservoir qubits'' to ``ancillary qubits''. Only the latter are targeted to extract classical information, with minimal disturbance to the reservoir quantum state. 
This strategy implies a careful trade-off between the strength of the measurement, the memory capacity of the reservoir, and the computational overhead of resetting the system to successfully perform on benchmark tasks~\cite{mujal:weak_measurements, franceschetto:weak_measurements}.

On the other side, a proper reservoir's behaviour shall prioritise recent inputs, and thus needs to erase information over time. Mathematically, it means that the map that describes the reservoir state evolution, $r$ in Eq.~\eqref{Eq:fading_memory}, needs to be a contraction. 
As a consequence, the fully unitary dynamics introduced in Sect.~\ref{sec:intro_QRC} is not suitable for QRC purposes. 
Again, in our hQCRC setting, this issue is overcome by post-processing the measured data from the quantum system with a classical fading memory function. 
However, moving to a fully quantum regime, fading memories could only be obtained if non-negligible losses are present \cite{Sannia2024dissipationas}. 
Within this strategy, an envisioned setting is to evolve an initial quantum state but extract classical information only via measurements on a subsystem.  
Concretely, for a neutral atom system compatible with the one in Fig.~\ref{fig:hQCRC_network}, one starts with an $N$-qubit initial state $|0\rangle^{\otimes N}$, and maps e.g. the transformed input $Y_1 (t)$ to a set of Hamiltonian parameters $Y_1^{(1,2,3)}$. 
Due to the presence of losses, the evolved state after a fixed time $t$ becomes a more complex function $\mathcal{L}(Y_1^{(1,2,3)},t)$ generalising over Eq.~\eqref{eq:qrc_state}, where $\mathcal{L}$ is the so-called \textit{Liouvillian} which models the lossy reservoir evolution.
The reservoir state $r_1$ is extracted by weak-measuring a subset of $M < N$ atoms, e.g., repeating the experiments so that finite-shot expectation values of local observables can be computed.  
For each subsequent input $u_t$, the procedure can be repeated, but the additional $\mathcal{L}(Y_n^{(1,2,3)},t)$ acts on the quantum state as evolved until the previous $n-1$ timestep, and altered by the weak measurement. 

In a more general case, depending on the particular encoding, different physical properties of the quantum system itself can be used for introducing effective losses to the system. For example, if the input is encoded in a reservoir subsystem, scrambling dynamics are favourable for having a fading memory~\cite{martinez:eth}. 
Also, many open quantum systems are described by a \textit{Markovian} behaviour, which erases information of past inputs exponentially fast over time~\cite{sannia2025nonmarkovianity}, precluding the possibility of solving tasks where a coexistence of short and long time correlations is needed. 
This challenge, in a fully quantum case, can only be addressed with the introduction of quantum non-Markovianity~\cite{sannia2025nonmarkovianity}. However, hybrid scenarios like the one introduced can leverage upon classical memory to allow for such longer time correlations.

Another fundamental challenge is the extraction of information from the quantum device, encompassing beyond RC to more generic quantum machine learning tasks. 
Data processing is expected to occur in the vast Hilbert space available already to modest-sized quantum devices, however it is often lower-dimensional classical data that are required as the usable output for real-world tasks. This throughput mismatch is known as the \textit{readout problem}, {and it is closely related to concentration phenomena}. 
{Specific techniques are being explored} to extract such crux of information \cite{williams2024addressing}, often at the heart of the performance of quantum algorithms. 
In our specific case, this readout challenge emerges naturally e.g. when adopting ancillary subsystems and scrambling dynamics, because the number of measurements needed to properly compute the reservoir response can exponentially scale with the reservoir dimension~\cite{xiong2025rolescramblingnoisetemporal, sannia2025exponential} and only the presence of appropriate symmetries can solve the issue \cite{sannia2025exponential}. 
A measurement overhead problem can also be present in the neutral atom setting considered here, as the system size scales and (linearly) more expectations values need being computed in Eq.~\eqref{eq:expvals}. Possible strategies for dealing with this issue will be the topic of further research.

On the positive side, QRC does not require expensive circuit gradient estimate and optimization~\cite{jaeger:echo}, which affects instead other variational QML techniques. Therefore, it could elude trainability issues connected with the emergence of complex or flat optimization landscapes~\cite{larocca2025barren}, whilst harnessing the expressivity accessible to complex, high-dimensional quantum states to potentially capture patterns in specific datasets better than classical tools~\cite{kornjavca2024qrc}. 

Finally, we can safely assume that noise limitations in the quantum device of choice for QRC implementations will play a crucial role in dictating architectures. 
At a stage in quantum computing where extremely different technologies coexist, to the point that appropriate cross-platform benchmarking is still an actively developed and debated research topic~\cite{lorenz2025systematic, proctor2025benchmarking}, it is expected how different quantum architectural choices - with their strengths and weaknesses in terms of gate fidelity, scalability in circuit width and depth, repetition rate and access to native operations - will influence the design of the corresponding QRC protocols. Indeed, each of these aspects will impact differently noise on the processing and readout phases.
In this contribution we focused on a Neutral Atom implementation, but even restricting ourselves to a single platform and its related noise model, we can envisage how a careful design and optimisation of the ansatz can bring massive performance benefits. The long-term complexity of this optimisation problem, as circuits scale beyond classical simulability and ease of result interpretation, will eventually invoke the aid of automated strategies, similarly to what has happened in the field of variational QML with Quantum Architecture Search schemes~\cite{martyniuk2024quantum}.

\vskip6pt

\section{Conclusions} 
We have presented a vision for quantum reservoir computing implemented in a hybrid quantum-classical framework (hQCRC). 
The purpose of this approach is to mitigate all shortcomings of a quantum system for being a potent reservoir, while taking advantage of exponentially large Hilbert spaces. 

In this regard, we showed that (quantum) feature maps represent an important tool for reservoir computing, as they enable processing in such high-dimensional latent space.
We mitigated the impact of destructive measurements by moving the memory part of the reservoir from the quantum subroutine to a classical one. 
We also combined the encoding of input data into quantum states through analog evolution (data-to-circuit), with non-linearities associated with the measurement protocol. 
Additionally, our hQCRC framework uses 
feature maps on the level of classical reservoir states, making it more flexible to accommodate complex tasks. 
The balance between purely classical and hybrid processing in our approach can be tuned and the quantum part easily turned completely off, to study the impact of the latter on task performances.

Crucially, we highlighted how our hQCRC can fit current and future hardware. We demonstrated how the mapping described above can invoke arrays of interacting qubits, where reconfigurable Rydberg neutral atom arrays represent a prominent example of such a quantum architecture.
The modularity of the quantum circuit construction enables us to scale complexity of each component in the framework - type and number of operations addressing the qubits, the size of the register and the number of extracted observables (which translates into reservoir sizes). 
Therefore, one can adjust the hQCRC approach to meet hardware restrictions. 

We demonstrated in a proof-of-concept experiment, how the hQCRC implemented on a simulated neutral-atom platform can predict future evolution of chaotic time-series system, and the quantum subroutine can be effectively exploited to augment the performance of a classical counterpart. 
Our results show how the efficient combination of quantum and classical resources can be beneficial for information processing in typical reservoir computing tasks, such as time-series prediction. Our work opens a route to new quantum-classical reservoir computing approaches and raises the question of useful resources in the hybrid domain.

\paragraph*{Acknowledgements}{C.G. O.K. \& A.A.G. acknowledge the support from the European Union's Horizon Europe research and innovation program under Grant Agreement No. 101070347-MANNGA. Views and opinions expressed are those of the authors only and do not necessarily reflect those of the European Union or the European Health and Digital Executive Agency (HADEA).  Neither the European Union nor the granting authority can be held responsible for them. 
O.K. acknowledges support from the QCi3 Hub (grant number EP/Z53318X/1). A. S. acknowledges the Spanish State Research Agency, through the Mar\'ia de Maeztu project CEX2021-001164-M funded by the MCIU/AEI/10.13039/501100011033, through the COQUSY project PID2022-140506NB-C21 and -C22 funded by MCIU/AEI/10.13039/501100011033, MINECO through the QUANTUM SPAIN project, and EU through the RTRP - NextGenerationEU within the framework of the Digital Spain 2025 Agenda. A.S. also acknowledges the CSIC Interdisciplinary Thematic Platform (PTI+) on Quantum Technologies in Spain (QTEP+) and the support of a fellowship from the ``la Caixa” Foundation (ID 100010434 - LCF/BQ/DI23/11990081).}


\bibliographystyle{unsrt}
\bibliography{biblio}

\begin{thebibliography}{10}

\bibitem{proctor2025benchmarking}
Timothy Proctor, Kevin Young, Andrew~D Baczewski, and Robin Blume-Kohout.
\newblock Benchmarking quantum computers.
\newblock {\em Nature Reviews Physics}, pages 1--14, 2025.

\bibitem{callison2022hybrid}
Adam Callison and Nicholas Chancellor.
\newblock Hybrid quantum-classical algorithms in the noisy intermediate-scale
  quantum era and beyond.
\newblock {\em Physical Review A}, 106(1):010101, 2022.

\bibitem{wudarski2023hybrid}
Filip Wudarski, Daniel OConnor, Shaun Geaney, Ata~Akbari Asanjan, Max Wilson,
  Elena Strbac, P~Aaron Lott, and Davide Venturelli.
\newblock Hybrid quantum-classical reservoir computing for simulating chaotic
  systems.
\newblock {\em preprint arXiv:2311.14105}, 2023.

\bibitem{platt2022systematic}
Jason~A Platt, Stephen~G Penny, Timothy~A Smith, Tse-Chun Chen, and Henry~DI
  Abarbanel.
\newblock A systematic exploration of reservoir computing for forecasting
  complex spatiotemporal dynamics.
\newblock {\em Neural Networks}, 153:530--552, 2022.

\bibitem{yan:qrc_overview}
Min Yan, Can Huang, Peter Bienstman, Peter Tino, Wei Lin, and Jie Sun.
\newblock Emerging opportunities and challenges for the future of reservoir
  computing.
\newblock {\em Nature Communications}, 15, 2024.

\bibitem{markovic2020neurophysics}
Danijela Markovi{\'c}, Alice Mizrahi, Damien Querlioz, and Julie Grollier.
\newblock Physics for neuromorphic computing.
\newblock {\em Nature Reviews Physics}, 2(9):499--510, 2020.

\bibitem{gauthier2021nextrc}
Daniel~J Gauthier, Erik Bollt, Aaron Griffith, and Wendson~AS Barbosa.
\newblock Next generation reservoir computing.
\newblock {\em Nature communications}, 12(1):5564, 2021.

\bibitem{huang2025vast}
Hsin-Yuan Huang, Soonwon Choi, Jarrod~R McClean, and John Preskill.
\newblock The vast world of quantum advantage.
\newblock {\em preprint arXiv:2508.05720}, 2025.

\bibitem{jaeger:memory}
Herbert Jaeger and Harald Haas.
\newblock Harnessing nonlinearity: Predicting chaotic systems and saving energy
  in wireless communication.
\newblock {\em Science}, 304(5667):78--80, 2004.

\bibitem{jaeger:echo}
Herbert Jaeger.
\newblock The “echo state” approach to analysing and training recurrent
  neural networks-with an erratum note.
\newblock {\em German NRC for information technology - technical report}, 148,
  2001.

\bibitem{gallicchio:echo_measure}
Claudio Gallicchio et~al.
\newblock Chasing the echo state property.
\newblock In {\em ESANN 2019-Proceedings, 27th European Symposium on Artificial
  Neural Networks, Computational Intelligence and Machine Learning}. ESANN,
  2019.

\bibitem{zhong:memristors}
Yanan Zhong, Jianshi Tang, Xinyi Li, Bin Gao, He~Qian, and Huaqiang Wu.
\newblock Dynamic memristor-based reservoir computing for high-efficiency
  temporal signal processing.
\newblock {\em Nature communications}, 12, 2021.

\bibitem{finocchio:spintronics}
Giovanni Finocchio, Massimiliano Di~Ventra, Kerem~Y Camsari, Karin
  Everschor-Sitte, Pedram~Khalili Amiri, and Zhongming Zeng.
\newblock The promise of spintronics for unconventional computing.
\newblock {\em Journal of Magnetism and Magnetic Materials}, 521, 2021.

\bibitem{van:photonics}
Guy Van~der Sande, Daniel Brunner, and Miguel~C Soriano.
\newblock Advances in photonic reservoir computing.
\newblock {\em Nanophotonics}, 6, 2017.

\bibitem{penkovsky:FPGA}
Bogdan Penkovsky, Laurent Larger, and Daniel Brunner.
\newblock Efficient design of hardware-enabled reservoir computing in fpgas.
\newblock {\em Journal of Applied Physics}, 124, 2018.

\bibitem{fujii:qrc_og}
Keisuke Fujii and Kohei Nakajima.
\newblock Quantum reservoir computing: a reservoir approach toward quantum
  machine learning on near-term quantum devices.
\newblock {\em Reservoir Computing: Theory, Physical Implementations, and
  Applications}, 2021.

\bibitem{markovic:qrc_og}
Danijela Markovi{\'c} and Julie Grollier.
\newblock Quantum neuromorphic computing.
\newblock {\em Applied physics letters}, 2020.

\bibitem{mujal:qrc_overview}
Pere Mujal, Rodrigo Mart{\'\i}nez-Pe{\~n}a, Johannes Nokkala, Jorge
  Garc{\'\i}a-Beni, Gian~Luca Giorgi, Miguel~C Soriano, and Roberta Zambrini.
\newblock Opportunities in quantum reservoir computing and extreme learning
  machines.
\newblock {\em Advanced Quantum Technologies}, 4, 2021.

\bibitem{albrecht2023quantum}
Boris Albrecht, Constantin Dalyac, Lucas Leclerc, et~al.
\newblock Quantum feature maps for graph machine learning on a neutral atom
  quantum processor.
\newblock {\em Physical Review A}, 107(4):042615, 2023.

\bibitem{schuld2019quantum}
Maria Schuld and Nathan Killoran.
\newblock Quantum machine learning in feature hilbert spaces.
\newblock {\em Physical review letters}, 122(4):040504, 2019.

\bibitem{kyriienko2021solving}
Oleksandr Kyriienko, Annie~E Paine, and Vincent~E Elfving.
\newblock Solving nonlinear differential equations with differentiable quantum
  circuits.
\newblock {\em Physical Review A}, 103(5):052416, 2021.

\bibitem{henriet:pasqal}
Lo{\"\i}c Henriet, Lucas Beguin, Adrien Signoles, Thierry Lahaye, Antoine
  Browaeys, Georges-Olivier Reymond, and Christophe Jurczak.
\newblock Quantum computing with neutral atoms.
\newblock {\em Quantum}, 4, 2020.

\bibitem{bluvstein2024logical}
Dolev Bluvstein, Simon~J Evered, Alexandra~A Geim, Sophie~H Li, Hengyun Zhou,
  Tom Manovitz, Sepehr Ebadi, Madelyn Cain, Marcin Kalinowski, Dominik
  Hangleiter, et~al.
\newblock Logical quantum processor based on reconfigurable atom arrays.
\newblock {\em Nature}, 626(7997):58--65, 2024.

\bibitem{labuhn2015realizing}
Henning Labuhn, Daniel Barredo, Sylvain Ravets, Sylvain de~L{\'e}s{\'e}leuc,
  Tommaso Macr{\`\i}, Thierry Lahaye, and Antoine Browaeys.
\newblock Realizing quantum ising models in tunable two-dimensional arrays of
  single rydberg atoms.
\newblock {\em preprint arXiv:1509.04543}, 2015.

\bibitem{kornjavca2024qrc}
Milan Kornja{\v{c}}a, Hong-Ye Hu, Chen Zhao, Jonathan Wurtz, Phillip Weinberg,
  Majd Hamdan, Andrii Zhdanov, Sergio~H Cantu, Hengyun Zhou, Rodrigo~Araiza
  Bravo, et~al.
\newblock Large-scale quantum reservoir learning with an analog quantum
  computer.
\newblock {\em preprint arXiv:2407.02553}, 2024.

\bibitem{settino2025memory}
J~Settino, L~Salatino, L~Mariani, F~D’Amore, M~Channab, L~Bozzolo, S~Vallisa,
  P~Barill{\`a}, A~Policicchio, N~Lo~Gullo, et~al.
\newblock Memory-augmented hybrid quantum reservoir computing.
\newblock {\em Physical Review Applied}, 24(2):024019, 2025.

\bibitem{parra2020digital}
Adrian Parra-Rodriguez, Pavel Lougovski, Lucas Lamata, Enrique Solano, and
  Mikel Sanz.
\newblock Digital-analog quantum computation.
\newblock {\em Physical Review A}, 101(2):022305, 2020.

\bibitem{silverio2022pulser}
Henrique Silv{\'e}rio, Sebasti{\'a}n Grijalva, Constantin Dalyac, Lucas
  Leclerc, Peter~J Karalekas, Nathan Shammah, Mourad Beji, Louis-Paul Henry,
  and Lo{\"\i}c Henriet.
\newblock Pulser: An open-source package for the design of pulse sequences in
  programmable neutral-atom arrays.
\newblock {\em Quantum}, 6:629, 2022.

\bibitem{pasqalcloud}
Pasqal SaS~Cloud Services, 2025.
\newblock https://docs.pasqal.com/cloud/.

\bibitem{lukovsevivcius2012practical}
Mantas Luko{\v{s}}evi{\v{c}}ius.
\newblock A practical guide to applying echo state networks.
\newblock In {\em Neural Networks: Tricks of the Trade: Second Edition}, pages
  659--686. Springer, 2012.

\bibitem{lorenz1963deterministic}
Edward~N Lorenz.
\newblock Deterministic nonperiodic flow.
\newblock {\em Journal of atmospheric sciences}, 20(2):130--141, 1963.

\bibitem{zenodo}
Zenodo~Public Repository.
\newblock 10.5281/zenodo.17153722, 2025.
\newblock Code in this repo are embargoed until 1/9/2026. Please contact the
  Authors for early access.

\bibitem{mujal:weak_measurements}
Pere Mujal, Rodrigo Mart{\'\i}nez-Pe{\~n}a, Gian~Luca Giorgi, Miguel~C Soriano,
  and Roberta Zambrini.
\newblock Time-series quantum reservoir computing with weak and projective
  measurements.
\newblock {\em npj Quantum Information}, 9, 2023.

\bibitem{franceschetto:weak_measurements}
Giacomo Franceschetto, Marcin P{\l}odzie{\'n}, Maciej Lewenstein, Antonio
  Ac{\'\i}n, and Pere Mujal.
\newblock Harnessing quantum back-action for time-series processing.
\newblock {\em preprint arXiv:2411.03979}, 2024.

\bibitem{Sannia2024dissipationas}
Antonio Sannia, Rodrigo Mart{\'{i}}nez-Pe{\~{n}}a, Miguel~C. Soriano, Gian~Luca
  Giorgi, and Roberta Zambrini.
\newblock Dissipation as a resource for {Q}uantum {R}eservoir {C}omputing.
\newblock {\em {Quantum}}, 8:1291, March 2024.

\bibitem{martinez:eth}
Rodrigo Mart{\'\i}nez-Pe{\~n}a, Gian~Luca Giorgi, Johannes Nokkala, Miguel~C
  Soriano, and Roberta Zambrini.
\newblock Dynamical phase transitions in quantum reservoir computing.
\newblock {\em Physical Review Letters}, 127, 2021.

\bibitem{sannia2025nonmarkovianity}
Antonio Sannia, Ricard~Ravell Rodríguez, Gian~Luca Giorgi, and Roberta
  Zambrini.
\newblock Non-markovianity and memory enhancement in quantum reservoir
  computing.
\newblock {\em preprint arXiv:2505.02491}, 2025.

\bibitem{williams2024addressing}
Chelsea~A Williams, Stefano Scali, Antonio~A Gentile, Daniel Berger, and
  Oleksandr Kyriienko.
\newblock Addressing the readout problem in quantum differential equation
  algorithms with quantum scientific machine learning.
\newblock {\em preprint arXiv:2411.14259}, 2024.

\bibitem{xiong2025rolescramblingnoisetemporal}
Weijie Xiong, Zoë Holmes, Armando Angrisani, Yudai Suzuki, Thiparat Chotibut,
  and Supanut Thanasilp.
\newblock Role of scrambling and noise in temporal information processing with
  quantum systems.
\newblock {\em preprint arXiv:2505.10080}, 2025.

\bibitem{sannia2025exponential}
Antonio Sannia, Gian~Luca Giorgi, and Roberta Zambrini.
\newblock Exponential concentration and symmetries in quantum reservoir
  computing.
\newblock {\em preprint arXiv:2505.10062}, 2025.

\bibitem{larocca2025barren}
Mart{\'\i}n Larocca, Supanut Thanasilp, Samson Wang, Kunal Sharma, Jacob
  Biamonte, Patrick~J Coles, Lukasz Cincio, Jarrod~R McClean, Zo{\"e} Holmes,
  and M~Cerezo.
\newblock Barren plateaus in variational quantum computing.
\newblock {\em Nature Reviews Physics}, pages 1--16, 2025.

\bibitem{lorenz2025systematic}
Jeanette~Miriam Lorenz, Thomas Monz, Jens Eisert, Daniel Reitzner, F{\'e}licien
  Schopfer, Fr{\'e}d{\'e}ric Barbaresco, Krzysztof Kurowski, Ward van~der
  Schoot, Thomas Strohm, Jean Senellart, et~al.
\newblock Systematic benchmarking of quantum computers: status and
  recommendations.
\newblock {\em preprint arXiv:2503.04905}, 2025.

\bibitem{martyniuk2024quantum}
Darya Martyniuk, Johannes Jung, and Adrian Paschke.
\newblock Quantum architecture search: a survey.
\newblock In {\em 2024 IEEE International Conference on Quantum Computing and
  Engineering (QCE)}, volume~1, pages 1695--1706. IEEE, 2024.

\end{thebibliography}

\end{document}